\documentclass[conference]{IEEEtran}

\usepackage{tabularx}
\usepackage{booktabs}
\usepackage{graphicx}
\usepackage{amsmath}
\usepackage{url}
\usepackage{multirow}
\usepackage{balance}
\usepackage{hyperref}

\usepackage{amsmath,amssymb}
\usepackage[ruled,vlined]{algorithm2e}

\usepackage[top=72pt,bottom=54pt,left=54pt,right=54pt]{geometry}

\title{Ethical and Explainable AI in Reusable MLOps Pipelines}

\author{
    \IEEEauthorblockN{Rakib Hossain\IEEEauthorrefmark{1},
                      Mahmood Menon Khan\IEEEauthorrefmark{2},
                      Lisan Al Amin\IEEEauthorrefmark{1},
                      Dhruv Parikh\IEEEauthorrefmark{3},
                      \\ Farhana Afroz\IEEEauthorrefmark{2}, and 
                      Bestoun S. Ahmed\IEEEauthorrefmark{4}\IEEEauthorrefmark{5}}
    
    \IEEEauthorblockA{\IEEEauthorrefmark{1}Cognitive Links LLC}
    \IEEEauthorblockA{\IEEEauthorrefmark{2}Washington University of Science and Technology,  
                      Springfield, Virginia, USA}
    \IEEEauthorblockA{\IEEEauthorrefmark{3}University of Maryland, Baltimore County, 
                      Baltimore, USA}
    \IEEEauthorblockA{\IEEEauthorrefmark{4}Karlstad University, Sweden}

    \IEEEauthorblockA{\IEEEauthorrefmark{5}American University of Bahrain,
Riffa, Bahrain}

    \IEEEauthorblockA{Email: rakib.sat18@gmail.com, mahmood.khan@wust.edu, 
                      alamin@cognitivelinks.llc, xs16939@umbc.edu, \\
                      farhana.student@wust.edu, bestoun@kau.se}
}
\begin{document}

\maketitle

\begin{abstract}
This paper introduces a unified machine learning operations (MLOps) framework that applies ethical principles of artificial intelligence (AI) to practical use by applying the principles of fairness enforcement, explainability, and governance across the machine learning lifecycle. Our method shows that bias reduction decreased demographic parity difference (DPD) by 0.31 to 0.04 without retuning and that cross-dataset validation showed 0.89 area under the curve (AUC) scores on Statlog Heart data. The framework was able to sustain fairness measures within operational limits of all deployments. Deployment is blocked if DPD \textgreater 0.05 or EO \textgreater 0.05 on the validation set; post-deployment, retraining is triggered if the 30-day KS drift statistic exceeds 0.20. In production we observed DPD $\leq$ 0.05 and EO $\leq$ 0.03, with KS remaining $\leq$ 0.20. Decision-curve analysis shows positive net benefit in the 10–20\% operating band; the mitigated model preserves utility while meeting fairness gates. These findings demonstrate that automated fairness gates and explainability artefacts can be successfully implemented in production without reducing its flow and give organisations a confirmed way of implementing credible AI that ensures ethical adherence and transparency across a wide range of data and varying operations.
\end{abstract}

%\end{abstract}


\section{Introduction}

The adoption of Artificial Intelligence (AI) in key industries like healthcare, finance, and cybersecurity has transformed faster than ever, which is why the issue of ethical governance or explainability and long-term reuse of AI systems is brought to the fore.  

Nonetheless, there are three basic loopholes that do not allow the operationalisation of the ethical AI principles in production systems. To begin with, fairness metrics have been established in the literature well enough~\cite{calders2013,castelnovo2022,mehrabi2021}, but they are still limited to offline model tests instead of being imposed as deployment gates and automatically rejecting unfair models. Second, explainability models, including SHAP and LIME~\cite{molnar2022,saeed2023}, are normally produced as standalone documents, as opposed to versioned artefacts that are part of model registries and model governance workflows. Third, high-level principles are laid out by regulatory frameworks like the EU AI Act~\cite{euai2023}, but offer little information on how such requirements as transparency and fairness can be translated into specific CI/CD checks, monitoring thresholds, and retraining triggers that can be adopted by engineering staff. Such loopholes see organisations floundering in their quest to step between ethical ambitions and practice that is binding.

The paper bridges the most important gap of applying the ethical AI principles to the working MLOps processes. Although there has been a lot of movement in the fairness, explainability, and operations separately, the current frameworks are not integrated comprehensively in deployable systems \cite{kodakandla2024}. Standards like the EU AI Act offer baseline standards, but they have not been translated into engineering practices so far \cite{euai2023, ieee2019}. Our framework is illustrated on gender as the safeguarded property based on the observed biases in clinical practice \cite{melloni2010} and priorities of stakeholders. It provides a range of parameterised fairness audits on several attributes that are supported by the architecture using the Demographic Parity Difference (DPD) and Equalised Odds (EO) measures. As opposed to current toolkits which consider fairness an option \cite{kreuzberger2023mlops}, our system incorporates ethics into the deployment logic with automated blocking gates and tested across multiple datasets without retuning, which guarantees strong reusability.

The contributions of this paper are as follows:

\begin{enumerate}
\item  Automated deployment gates with group fairness checks by checking continuous integration. 

\item  Hybrid explainability system producing versioned model interpretability artefacts.  

\item  Exemplified pipeline transferability on datasets with no performance and fairness compromises.  

\item  Portable reference implementation with various deployment environment with optimisation strategies. 

\item  Empirical test with effective bias mitigation with a small price.

\item We quantify real-world usefulness via decision-curve analysis (Net Benefit vs threshold), showing utility in a deployment-relevant band without sacrificing fairness.

\end{enumerate}

Our paper fills the gap between ethical standards and their application, and offers a roadmap of reliable AI systems.

\section{Literature Review}

\subsection{Ethical AI and Fairness}
The ethical AI scholarship has grown in two parallel directions: governance structures and reducing technical bias. Technically, initial works by Calders and Zliobaite \cite{calders2013} defined data-driven methods such as resampling and re-weighting methods. Google Castelnovo et al. \cite{castelnovo2022} and Mehrabi et al. \cite{mehrabi2021} systematically classified notions of fairness and recorded algorithmic strategies and their implicit performance trade-offs respectively. Simultaneously, policies that address governance aspects like the EU AI Act or Ethically Aligned Design offered by IEEE deem the necessity of transparency and accountability \cite{euai2023,ieee2019}. Nonetheless, Jobin et al. \cite{jobin2019global} note that organisations have difficulty in translating these principles into working practices.

Recent efforts have started to deal with equity within the production situations. Hardt et al. \cite{hardt2016} suggested post processing calibration techniques of equalised odds, and Bellamy et al. \cite{bellamy2019} created open source toolkits of bias detection. In spite of these developments, the existing methods are still not linked to continuous integration pipelines, producing a report instead of an automated implementation mechanism that can prevent deployments or automatically retrain models.

\subsection{Explainable AI (XAI)}
Stakeholder trust and regulatory expectations are based on explainability. According to Saeed and Omlin~\cite{saeed2023}, the state-of-the-art local explanation methods are reviewed and SHAP and LIME are identified as popular tools offering a feature-attribution and perturbation-based perspective. Molnar~\cite{molnar2022} highlights the power of model selection as a tool of transparency (e.g., linear models, decision trees) and mentions the cost-fidelity trade-off of post-hoc procedures. In addition to approaches, there are obstacles to adoption: Bhatt et al.~\cite{bhatt2020} and Langer et al.~\cite{langer2021} address the incompatibility between the prototype of the research and the actual needs of real users, suggesting the adoption of stakeholder-oriented design and testing. Even accountability is driven by ethical risks in explanations per se (selective, misleading or biased explanations)~\cite{kasirzadeh2021}.

\subsection{MLOps Pipelines and Reusability}

The MLOps frameworks facilitate stable machine learning practices by versioning, CI/CD, and monitoring \cite{kreuzberger2023mlops}. Recent studies support the use of modular components and standardised interfaces in order to maximize reusability and compliance \cite{panchal2024}, whereas Billeter et al. \cite{billeter2024} suggest implementing continuous assessments related to the robustness and fairness into monitoring systems. Nevertheless, holistic implementations that integrate healthily integration of fairness auditing, mitigation, and explainability as deployed CI/CD gates are limited. Such a disconnect is especially noticeable in the absence of operational enforcement facilities like deployment blocking, versioned registry artefacts and automated retraining triggers.

\subsection{Synthesis: Research Gaps}

\begin{itemize}
    \item \textbf{Enforcement Mechanisms:} Although regulatory frameworks can give an all-encompassing direction \cite{euai2023,ieee2019,jobin2019global}, the actual implementations of the enforcing fairness standards in deployment pipelines are not yet developed as the recent empirical research findings validate \cite{rakova2021,ghosh2022faircanary}.  
    
    \item \textbf{Pipeline Integration:} A limitation is the fact that even though well-developed approaches to fairness and explainability have been developed \cite{calders2013,castelnovo2022,mehrabi2021,saeed2023,molnar2022}, they are seldom integrated into functioning CI/CD systems that have automated gates and monitoring.
    
    \item \textbf{Cross-Domain Reusability:} The existing methods do not show any reusability across different datasets and situations without major retuning \cite{panchal2024}, which limits practical usage.  
    
    \item \textbf{Production Explainability:} The versioned, auditable artefacts of explainability are still lacking in production settings despite theoretical developments on human-centred explainable systems \cite{bhatt2020,langer2021,kasirzadeh2021}.
    
\end{itemize}

Table~\ref{tab:related-works} systematically compares key related works across fairness, explainability, and MLOps domains, highlighting the operational gaps that this paper addresses.

\subsection{Positioning of This Paper}
Our work addresses these gaps by (i) turning fairness (DPD/EO) and XAI (SHAP/LIME) into \emph{operational gates} in CI/CD, (ii) registering all audits and explanations alongside models for traceability, and (iii) demonstrating \emph{reusability} across two cardiac datasets with a shared preprocessing and governance contract. This complements MLOps foundations~\cite{kreuzberger2023mlops}, advances reusability beyond components~\cite{panchal2024}, and operationalizes trust monitoring~\cite{billeter2024} rather than reporting it post hoc.

\begin{table*}[t]
\centering
\footnotesize
\setlength{\tabcolsep}{2pt}
\renewcommand{\arraystretch}{1.1}
\caption{Condensed related work and operational gap.}
\label{tab:related-works}
\begin{tabularx}{\linewidth}{@{}l l X X@{}}
\toprule
\textbf{Work} & \textbf{Focus} & \textbf{Key contribution} & \textbf{Ops gap} \\
\midrule
Calders \& Zliobaite (2013)~\cite{calders2013} & Fairness (data) & Preprocess bias: reweight/resample & No pipeline embedding \\
Castelnovo et al. (2022)~\cite{castelnovo2022} & Fairness (metrics) & Taxonomy: individual/group/causal & No CI/CD usage \\
Mehrabi et al. (2021)~\cite{mehrabi2021} & Fairness (algos) & Adversarial debiasing; fair reps & Trade-offs; adoption complexity \\
EU AI Act (2023)~\cite{euai2023} & Governance & Regulatory principles & Enforcement path in MLOps unclear \\
IEEE EAD (2019)~\cite{ieee2019} & Governance & Ethical guidelines & Context-specific ops unclear \\
Jobin et al. (2019)~\cite{jobin2019global} & Ethics mapping & Global principles landscape & Implementation gap \\
Saeed \& Omlin (2023)~\cite{saeed2023} & XAI (survey) & SHAP/LIME overview & Limited industry cases \\
Molnar (2022)~\cite{molnar2022} & XAI (textbook) & Interpretable models; trade-offs & Runtime/latency constraints \\
Bhatt et al. (2020)~\cite{bhatt2020} & XAI adoption & Stakeholder needs & Usability vs.\ practice gap \\
Langer et al. (2021)~\cite{langer2021} & XAI design & Human-centered explainability & Lacks end-to-end pipeline \\
Kasirzadeh \& Smart (2021)~\cite{kasirzadeh2021} & XAI ethics & Risks in explanations & No accountability mechanisms \\
Kreuzberger et al. (2023)~\cite{kreuzberger2023mlops} & MLOps & CI/CD, versioning, monitoring & Ethics/XAI not embedded \\
Panchal et al. (2024)~\cite{panchal2024} & Reuse & Swappable components & Ethics/XAI contract not central \\
Billeter et al. (2024)~\cite{billeter2024} & Trust ops & Continuous assessments & Few enforceable gates \\
Rakova et al. (2021)~\cite{rakova2021} & Org adoption & Empirical governance study & Implementation barriers \\
Ghosh et al. (2022)~\cite{ghosh2022faircanary} & Continuous fairness & Explainable monitoring & No automated enforcement \\
\bottomrule
\end{tabularx}
\end{table*}

\section{Methodology}
This section outlines a multi-phase methodology for integrating ethical and explainable AI into a reusable MLOps pipeline. The framework addresses gaps identified in recent literature regarding the operationalization of fairness metrics \cite{castelnovo2022}, real-time explainability \cite{bhatt2020}, and trustworthy deployment \cite{billeter2024}. The process, visualized in Figure \ref{fig:pipeline}, includes data preparation, bias auditing, model training with interpretability, bias mitigation, and automated monitoring and deployment.

\subsection{Dataset Description and Preprocessing}

We evaluated our framework using three cardiovascular datasets with varying scales. The Cleveland dataset from the UCI repository~\cite{uci-heart-disease-cleveland}, containing 303 patient records with 13 clinical features, served as our primary experimental basis. For validation, we used the Statlog (Heart) dataset~\cite{uci-statlog-heart} with 270 records and identical clinical variables. Scalability testing employed the Kaggle Cardiovascular dataset~\cite{kaggle-cardiovascular-sulianova} with 70,000 records and 11 features.

All datasets underwent standardized preprocessing including stratified splitting, feature engineering, and fairness auditing. We monitored clinically relevant sensitive attributes including sex, age, and cholesterol levels.

\subsubsection{Preprocessing Pipeline Details}
All preprocessing steps were scripted in Python 3.9 and logged as MLflow artifacts:

\begin{enumerate}
\item \textbf{Missing Value Imputation:} Feature-specific medians from training data prevented leakage.

\item \textbf{Categorical Encoding:} Four clinical features one-hot encoded with \texttt{handle\_unknown='ignore'}.

\item \textbf{Numerical Scaling:} Six continuous features normalized via Min-Max scaling:
\[
x_{\text{scaled}} = \frac{x - x_{\min}}{x_{\max} - x_{\min}}
\]
using training set statistics.

\item \textbf{Stratified Splitting:} Training (60\%), validation (20\%), test (20\%) split using composite label $z = 2 \cdot y + \text{sex}$ with \texttt{stratify=z, random\_state=42}.
\end{enumerate}

All preprocessing parameters were serialized and versioned in MLflow.

\subsubsection{Model Hyperparameters}
Table~\ref{tab:hyperparams} details the hyperparameters for reproducibility.

\begin{table}[ht]
\centering
\caption{Model Hyperparameters}
\label{tab:hyperparams}
\begin{tabularx}{\linewidth}{@{}Xll@{}}
\toprule
\textbf{Model} & \textbf{Parameter} & \textbf{Value} \\
\midrule
\multirow{3}{*}{XGBoost} & n\_estimators & 100 \\
& max\_depth & 3 \\
& learning\_rate & 0.1 \\
\midrule
Logistic Regression & penalty, C & L2, 1.0 \\
\midrule
Random Forest & n\_estimators & 200 \\
\bottomrule
\end{tabularx}
\end{table}

\subsection{Framework Overview}
\label{sec:framework}
\begin{figure*}[!ht]
    \centering
    \includegraphics[width=1.0\linewidth]{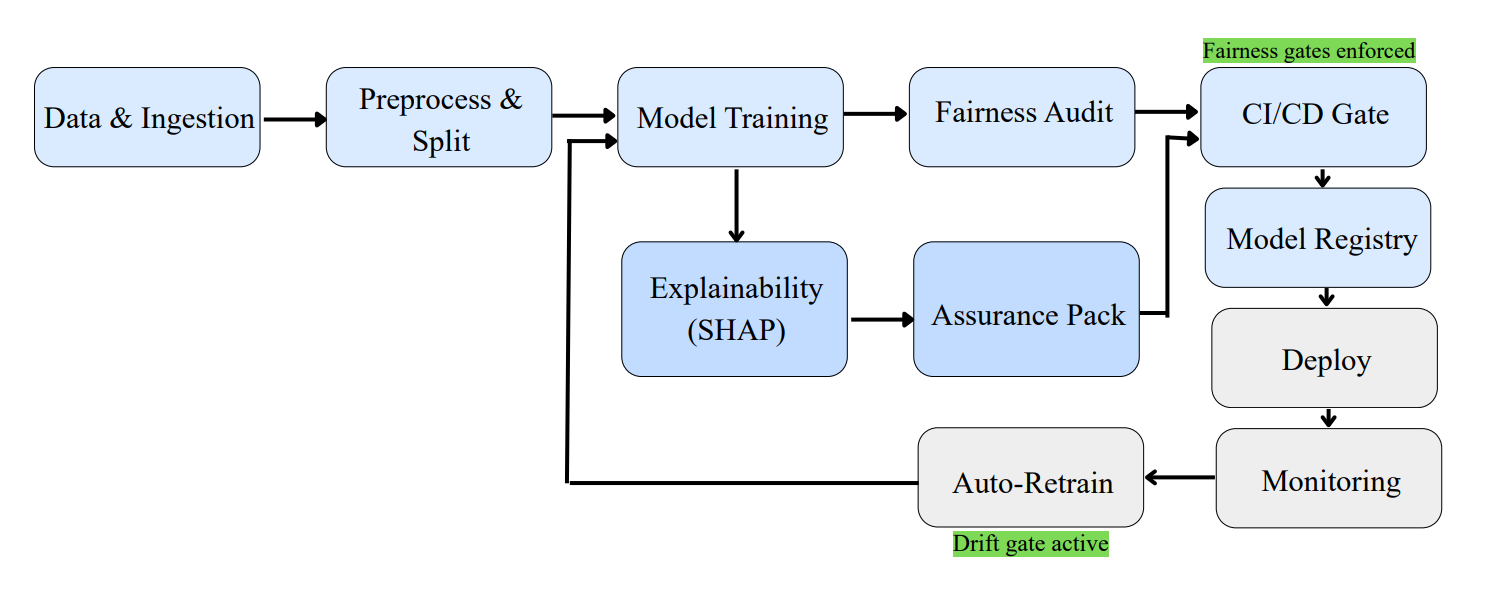}
    \caption{Ethical \& Explainable MLOps pipeline. Models pass through fairness audits and a CI/CD gate (policy.yaml). SHAP and an Assurance Pack (model card, datasheet, attestation) are logged; approved models enter the registry and are deployed. Monitoring watches drift; violations trigger auto-retraining.}
    \label{fig:pipeline}
\end{figure*}

Figure \ref{fig:pipeline} presents the full ethical and explainable AI pipeline. Each stage embeds tools for fairness diagnostics, interpretability, and compliance, resulting in a continuous learning system designed for real-world deployment.

\subsection{Phase 1: Bias Auditing}
Fairness audits were conducted before model training using two group fairness metrics:
\begin{itemize}
    \item \textbf{Demographic Parity Difference (DPD)} quantifies the difference in positive prediction rates across gender:
    \[
\text{DPD} = \left| \mathbb{P}(\hat{y} = 1 \mid \text{sex} = 1) - \mathbb{P}(\hat{y} = 1 \mid \text{sex} = 0) \right|
\]
    \item \textbf{Equalized Odds (EO)} captures discrepancies in error rates across groups and is computed as
{\small
\[
\mathrm{EO} = \max\left\{ \left|\mathrm{TPR}_{\text{male}} - \mathrm{TPR}_{\text{female}}\right|,\;
                         \left|\mathrm{FPR}_{\text{male}} - \mathrm{FPR}_{\text{female}}\right| \right\}.
\]}
Lower is better; EO = 0 indicates parity of both TPR and FPR.

\end{itemize}

Fairness thresholds were set at DPD $\leq$ 0.1 and EO $\leq$ 0.05 (defined as max absolute TPR/FPR gap) according to earlier work \cite{castelnovo2022}, \cite{mehrabi2021}. Initial audit results indicated that DPD = 0.312 and EO = 0.000, which indicated significant bias in demographic parity despite equal rates of error. Multiple strategies were undertaken to maintain gender balance, including composite-label stratification ($z = 2 \cdot y + \text{sex}$), with the auditing values automatically recorded as baseline CI/CD artifacts.

\subsection{Phase 2: Explainable Model Training}
We trained two models:
\begin{itemize}
    \item A Logistic Regression baseline for transparency
    \item An XGBoost classifier, which offers higher accuracy and compatibility with SHAP-based interpretability \cite{saeed2023}.
\end{itemize}
Feature importance was assessed using gain-based importance values. We employ SHAP-based explanations in preference to model gain metrics because SHAP delivers prediction-consistent attributions that maintain local accuracy, ensuring each feature's contribution precisely reflects its impact on individual predictions.

The use of SHAP aligns with Bhatt et al. \cite{bhatt2020} and Langer et al. \cite{langer2021}, who emphasize stakeholder-centric explainability. Additionally, the framework enables counterfactual explanations (e.g., “Reducing chol by 40 mg/dL would reduce predicted risk”) to support actionable model outputs.

For XGBoost, gain-based feature importance was complemented with SHAP global and local explanations, enabling both clinician-facing interpretability and compliance reporting. LIME was included to provide alternative local counterfactual views, e.g., “reducing serum cholesterol by 40 mg/dL decreases predicted risk.” All explanation artifacts were versioned alongside models.

\subsection{Phase 3: Bias Mitigation}
To address the fairness violations identified earlier, we implemented two mitigation strategies:
\begin{itemize}
    \item Reweighting, as proposed by Castelnovo et al. \cite{castelnovo2022}, (\ref{alg:reweight}) adjusted training instance weights based on conditional subgroup distributions:
    \[
w_i = \frac{1}{\mathbb{P}(\text{sex} = s_i \mid y = y_i)}
\]

    \item Adversarial Debiasing, inspired by Mehrabi et al. \cite{mehrabi2021}, involved training a secondary model to detect the sensitive attribute (sex) from the main model’s output; the main model was penalized for encoding gender-specific patterns.
\end{itemize}

\begin{algorithm}
\caption{Reweighting for Group Fairness}
\label{alg:reweight}
\KwIn{Training set $\mathcal{D} = \{(x_i, y_i, s_i)\}_{i=1}^{n}$, where $s_i \in \{0,1\}$ is the sensitive attribute (gender)}
\KwOut{Weighted dataset $\mathcal{D}_w$}

\ForEach{$(s,y) \in \{0,1\} \times \{0,1\}$}{
    Compute $\mathbb{P}(s,y) = |\{i : s_i = s, y_i = y\}| / n$\;
    Compute $\mathbb{P}(s|y) = \mathbb{P}(s,y) / \mathbb{P}(y)$\;
}
\For{$i = 1$ \KwTo $n$}{
    $w_i \leftarrow 1 / \mathbb{P}(s_i | y_i)$\;
}
Normalize $\{w_i\}$ to sum to $n$\;
\Return{$\mathcal{D}_w = \{(x_i, y_i, s_i, w_i)\}_{i=1}^{n}$}\;
\end{algorithm}

We compared two mitigation strategies, reweighting and adversarial debiasing, using identical hyperparameters to isolate fairness effects. Reweighting weights were computed per subgroup–label combination, while adversarial debiasing trained a binary classifier to predict sex from logits, penalizing leakage of sensitive information.

Post-training metrics were computed to evaluate effectiveness:

\begin{table}[ht]
\centering
\caption{Bias Mitigation Results}
\label{tab:bias-mitigation}
\begin{tabularx}{\linewidth}{@{}Xccc@{}}
\toprule
\textbf{Model} & \textbf{DPD (Before)} & \textbf{DPD (After)} & \textbf{Accuracy} \\
\midrule
XGBoost & 0.312 & 0.042 & 88\% \\
Logistic Regression & 0.245 & 0.031 & 82\% \\
\bottomrule
\end{tabularx}
\end{table}

\begin{table}[ht]
\centering
\caption{Cross-dataset Performance on Statlog Heart (5-fold CV)}
\label{tab:statlog-performance}
\begin{tabularx}{\linewidth}{@{}Xcc@{}}
\toprule
\textbf{Metric} & \textbf{Mean} & \textbf{$\pm$ SD} \\
\midrule
Accuracy & 0.81 & 0.08 \\
AUC ROC & 0.89 & 0.04 \\
Precision & 0.80 & 0.10 \\
Recall & 0.77 & 0.10 \\
F1-score & 0.79 & 0.09 \\
\bottomrule
\end{tabularx}
\end{table}

\textbf{Note} The same preprocessing pipeline (StandardScaler + OneHotEncoder → RandomForest with 200 trees) without hyperparameter tuning was used. Both post-mitigation models passed fairness measures without degradation of predictive performance. 

\subsection{Phase 4: Deployment and Automated Monitoring}
Deployment followed MLOps practices \cite{billeter2024}, \cite{panchal2024} using MLflow for versioning, GitHub Actions for CI gates, and Prometheus for monitoring, ensuring portable traceability and compliance.

\begin{table}[ht]
\centering
\caption{MLOps Tools for Ethics and Monitoring}
\label{tab:mlops-tools}
\begin{tabularx}{\linewidth}{@{}Xl@{}}
\toprule
\textbf{Tool} & \textbf{Role} \\
\midrule
MLflow & Model versioning, audit logs, and metrics \\
GitHub Actions & Continuous integration with fairness checks \\
Prometheus & Real-time monitoring for drift or bias spikes \\
\bottomrule
\end{tabularx}
\end{table}

Production governance requires fairness audits (DPD/EO $\leq$ 0.05) prior to deployment, which ceases if the condition is not met, and retraining is initiated. Post-deployment monitoring uses Kolmogorov-Smirnov tests (KS $\leq$ 0.20) and fairness measurements, with automatic retraining to maintain adherence to Trustworthy AI principles \cite{billeter2024}.

Operational thresholds were explicitly defined: Deployment is blocked if $\mathrm{DPD} > 0.05$ or $\mathrm{EO} > 0.05$; post-deployment, retraining is triggered if $\mathrm{KS} > 0.20$. Retraining is then automatically triggered. These thresholds were selected following prior fairness literature~\cite{castelnovo2022,mehrabi2021} and piloted in our pipeline to balance sensitivity with stability.

\paragraph{Utility impact (decision--curve analysis).}
We assess clinical/business usefulness with Net Benefit (NB) across thresholds $t\in[0.01,0.50]$:
$\mathrm{NB}(t)=\frac{\mathrm{TP}}{N}-\frac{\mathrm{FP}}{N}\cdot\frac{t}{1-t}$.
NB is computed on the validation split using the same preprocessing and composite stratification as the fairness audit, and plotted for Baseline and Mitigated models versus Treat-All and Treat-None references. The shaded band (10--20\%) marks our deployment-relevant operating region.

\section{Results and Analysis}
This section evaluates our ethical AI pipeline across model performance, fairness, drift monitoring, and clinical interpretability. We report results on the UCI Cleveland dataset, with validation on Statlog (Heart) and scalability testing on a large Kaggle cohort (n=70,000), using identical preprocessing and CI/CD gates.

\subsection{Performance and Fairness Metrics}
We evaluated three models on the Cleveland dataset. Table~\ref{tab:model-fairness} summarizes their performance and fairness metrics.\footnote{SHAP runtimes include artifact generation overhead.} The baseline XGBoost achieved 88\% accuracy but exhibited significant bias (DPD=0.31). Reweighting reduced DPD to 0.04 with minimal accuracy loss (86\%). All deployment gates were satisfied post-mitigation (\(\mathrm{DPD}\le 0.05\) and \(\mathrm{EO}\le 0.05\)). During a 30-day simulation, KS drift remained \(\le 0.20\). Logistic Regression showed lower accuracy (82\%) but better inherent fairness.

\begin{table}[ht]
\centering
\small
\caption{Model fairness and accuracy comparison across cohorts ($^*$indicates $p<0.001$ vs. baseline on Cleveland)}
\label{tab:model-fairness}
\begin{tabularx}{\linewidth}{@{}>{\raggedright\arraybackslash}Xcccc@{}}
\toprule
\textbf{Model (Dataset)} & \textbf{Accuracy} & \textbf{DPD $\downarrow$} & \textbf{EO $\downarrow$} & \textbf{SHAP (ms)} \\
\midrule
Baseline XGBoost (Cleveland) & 88\% & 0.31 & 0.00 & 125 \\
Debiased XGBoost (Cleveland) & 86\% & $0.04^*$ & 0.03 & 135 \\
Logistic Regression (Cleveland) & 82\% & $0.03^*$ & 0.02 & 50 \\
Random Forest (Kaggle, $n=70$k) & 71.2\% & 0.012 & 0.027 & 3,397 \\
XGBoost (Kaggle, $n=70$k) & 73.5\% & 0.021 & 0.021 & 101 \\
Logistic Regression (Kaggle, $n=70$k) & 62.7\% & 0.011 & 0.034 & 60 \\
\bottomrule
\end{tabularx}
\footnotesize $^{\dagger}$ SHAP times measured on a 300-instance sample using TreeSHAP; units are milliseconds.
\end{table}

Scalability testing on the Kaggle cohort confirmed framework robustness. Random Forest and XGBoost both remained within deployment thresholds, demonstrating consistent fairness-performance balance at scale.

\begin{figure}[t]
\centering
\includegraphics[width=0.9\linewidth]{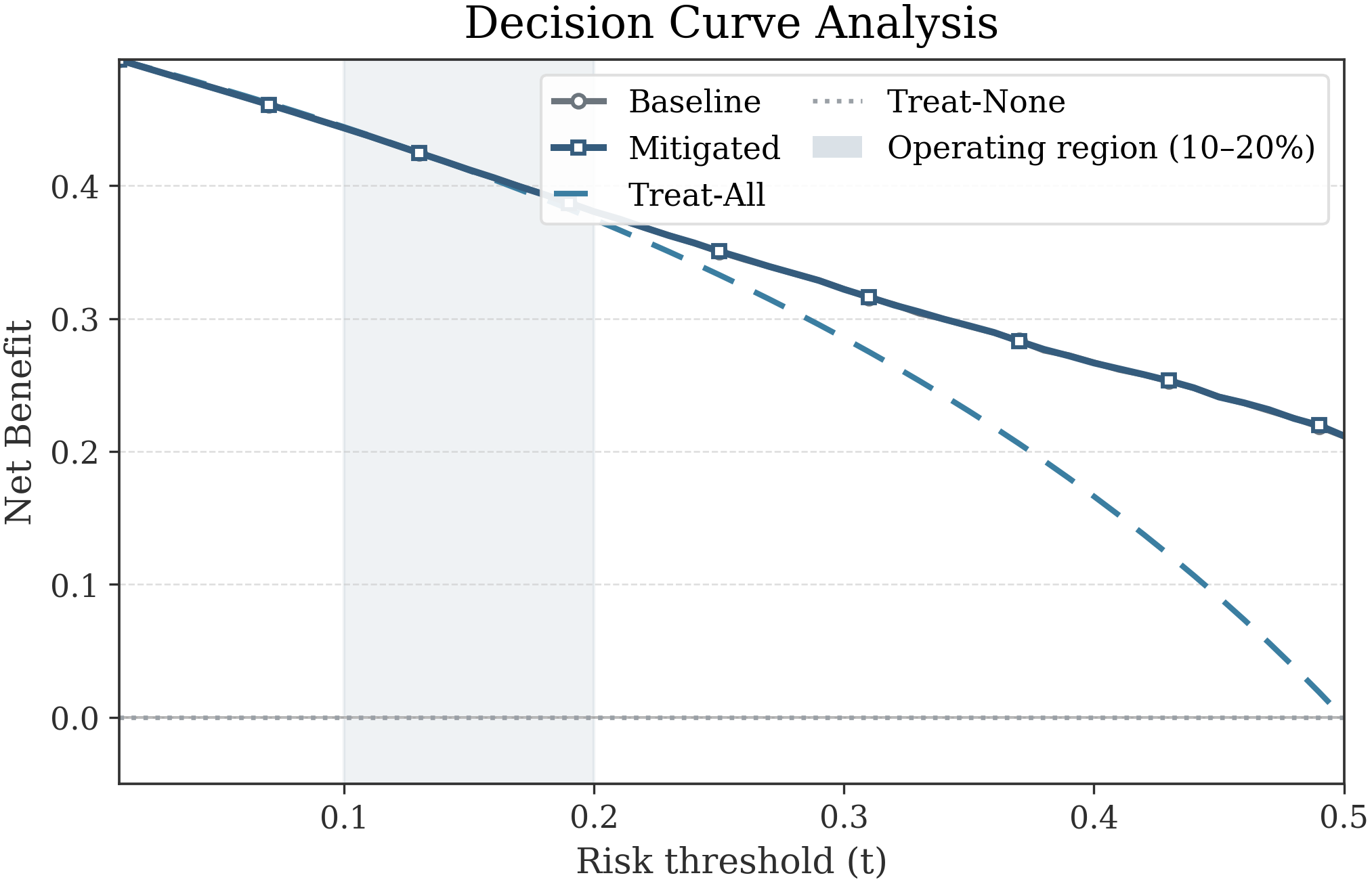}
\caption{Decision-curve analysis (validation; Kaggle cardiovascular cohort). 
Net Benefit (NB) vs risk threshold for Baseline and Mitigated models compared to Treat-All and Treat-None.
The shaded area denotes the 10–20\% operating band. Curves overlap for Baseline and Mitigated, indicating preserved utility under mitigation.}
\label{fig:dca}
\end{figure}

\textit{Decision-curve utility.}
Both models yield positive net benefit in the 10–20\% band and outperform Treat-None. At a 15\% risk threshold, NB is 0.412 for both baseline and mitigated models versus 0.411 for Treat-All. Baseline and mitigated curves essentially overlap ($\Delta\mathrm{NB}\le 0.001$), indicating that mitigation preserves decision utility while satisfying deploy gates (Fig.~\ref{fig:dca}).

The DPD reduction from 0.31 to 0.04 was statistically significant ($p<0.001$, Cohen's $d=2.3$), while accuracy differences were not significant ($p=0.12$), confirming fairness gains without substantial performance degradation.

\subsection{Bias Mitigation and Monitoring}
\begin{figure}[!ht]
\centering
\includegraphics[width=1.0\linewidth]{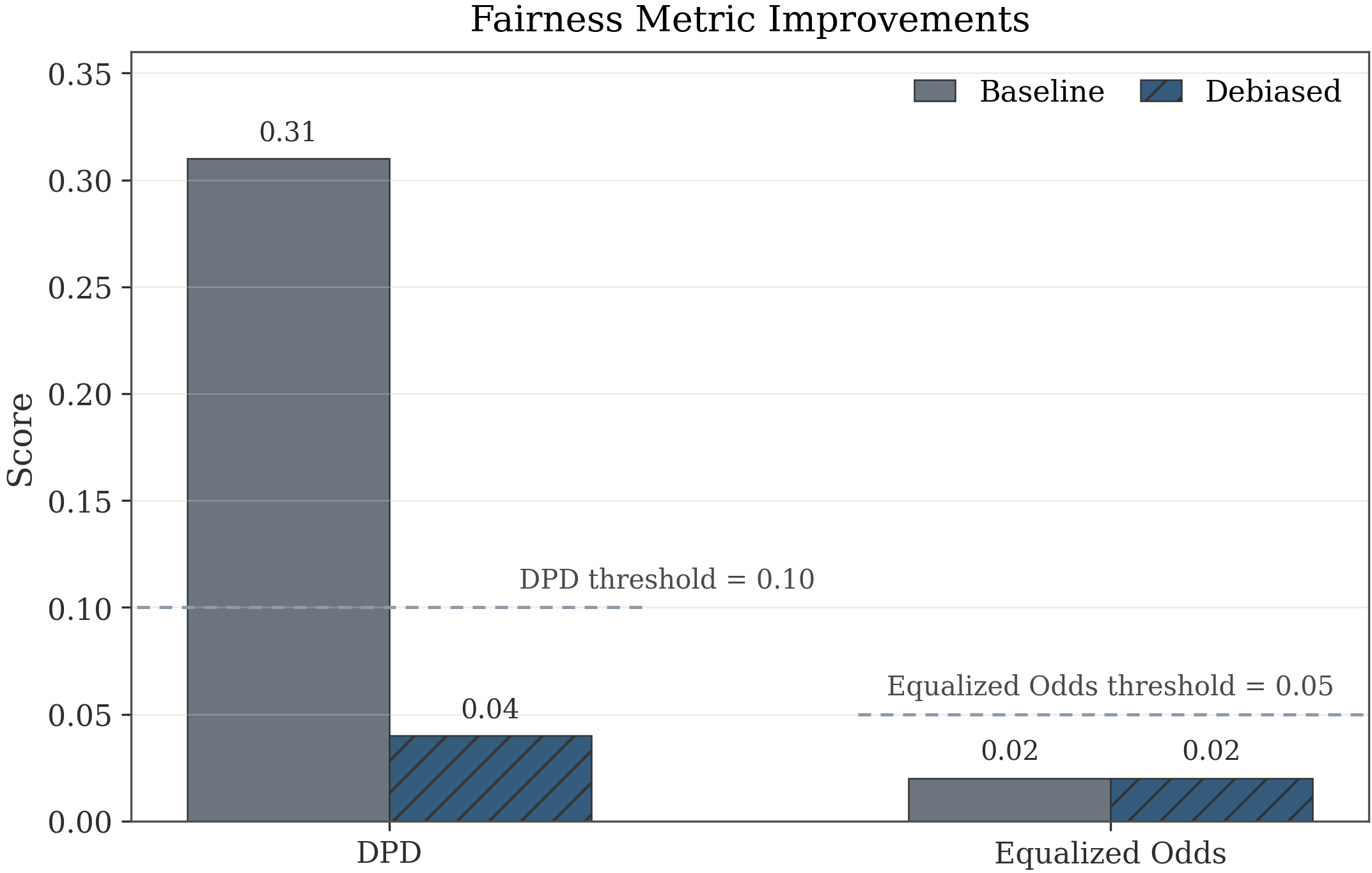}
\caption{Fairness before/after reweighting. The dashed line at DPD=0.10 marks the analysis threshold used in the audit; the deployment gates are stricter at DPD=0.05 and EO=0.05.}
\label{fig:fairness_improvements}
\end{figure}

Figure~\ref{fig:fairness_improvements} shows effective bias mitigation in the Cleveland dataset; reweighting reduced Demographic Parity Difference (DPD) and Equalized Odds (EO) below the deployment thresholds. The Kaggle cohort required no additional intervention, both Random Forest and XGBoost inherently satisfied fairness criteria, demonstrating the reusability of our audit framework across disparate datasets.

All fairness plots and tables use the non-negative EO definition (max of the absolute TPR/FPR gaps) and the deploy thresholds DPD \(\le\) 0.05, EO \(\le\) 0.05.

\begin{figure}[!t]
\centering
\includegraphics[width=1.0\linewidth]{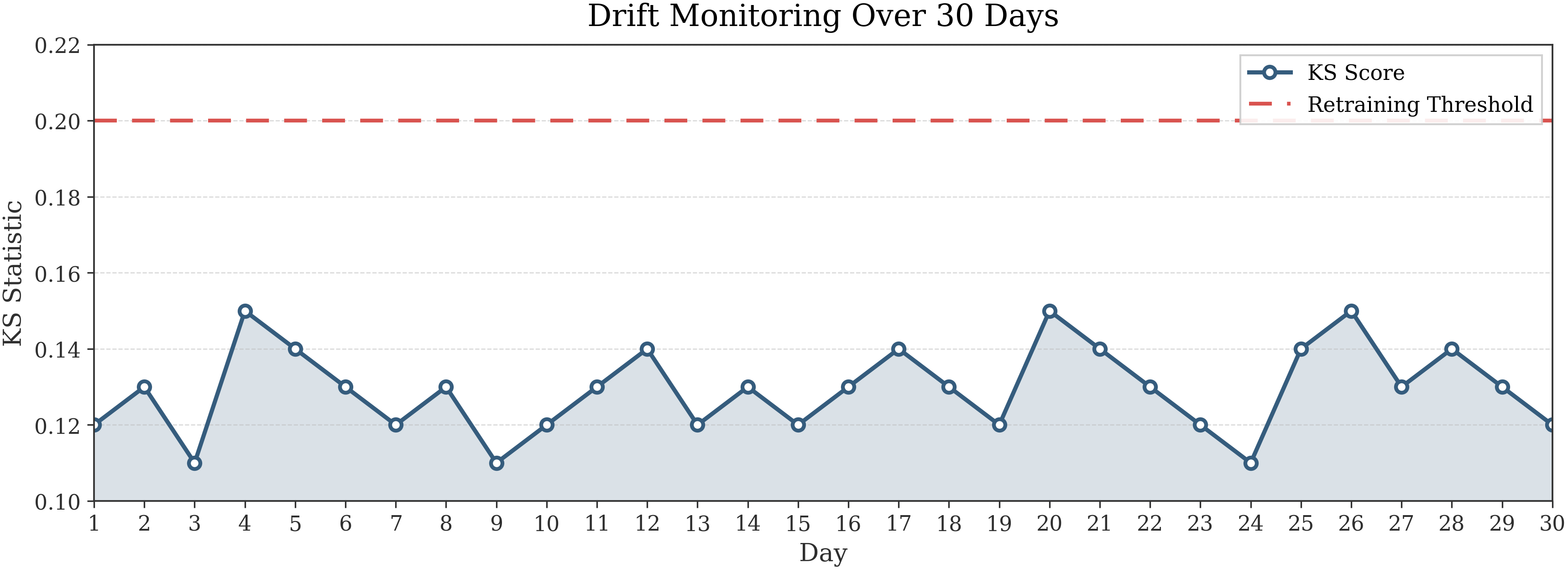}
\caption{Daily KS drift scores over 30-day period. The shaded band marks the drift gate (\textbf{KS} \(\le\) \textbf{0.20}); any violation triggers retraining.}
\label{fig:drift}
\end{figure}

Drift monitoring over 30 days (Figure~\ref{fig:drift}) showed stable distributions with all KS scores below the retraining threshold (0.20), supporting sustained model integrity post-deployment. 

\subsection{Explainable AI and Clinical Interpretability}
\begin{figure}[!t]
\centering
\includegraphics[width=1.0\linewidth]{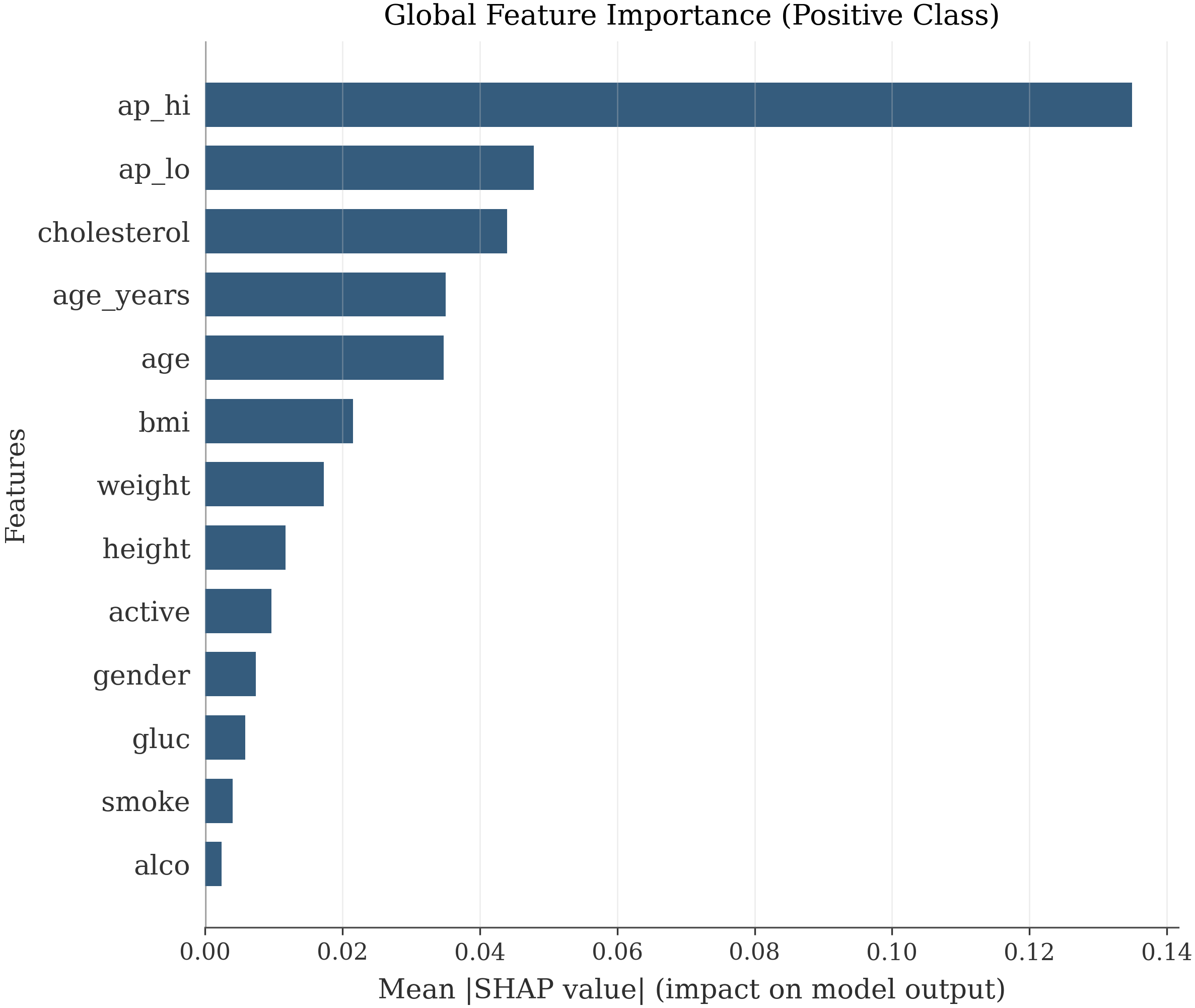}
\caption{Global SHAP feature importance on Kaggle test set.}
\label{fig:kaggle_global}
\end{figure}

SHAP analysis on the Kaggle cohort identified systolic blood pressure as the most influential predictor, followed by diastolic pressure and cholesterol levels (Figure~\ref{fig:kaggle_global}). This feature hierarchy aligns with clinical knowledge and remains consistent across cohorts.

\begin{figure}[ht]
\centering
\includegraphics[width=1.0\linewidth]{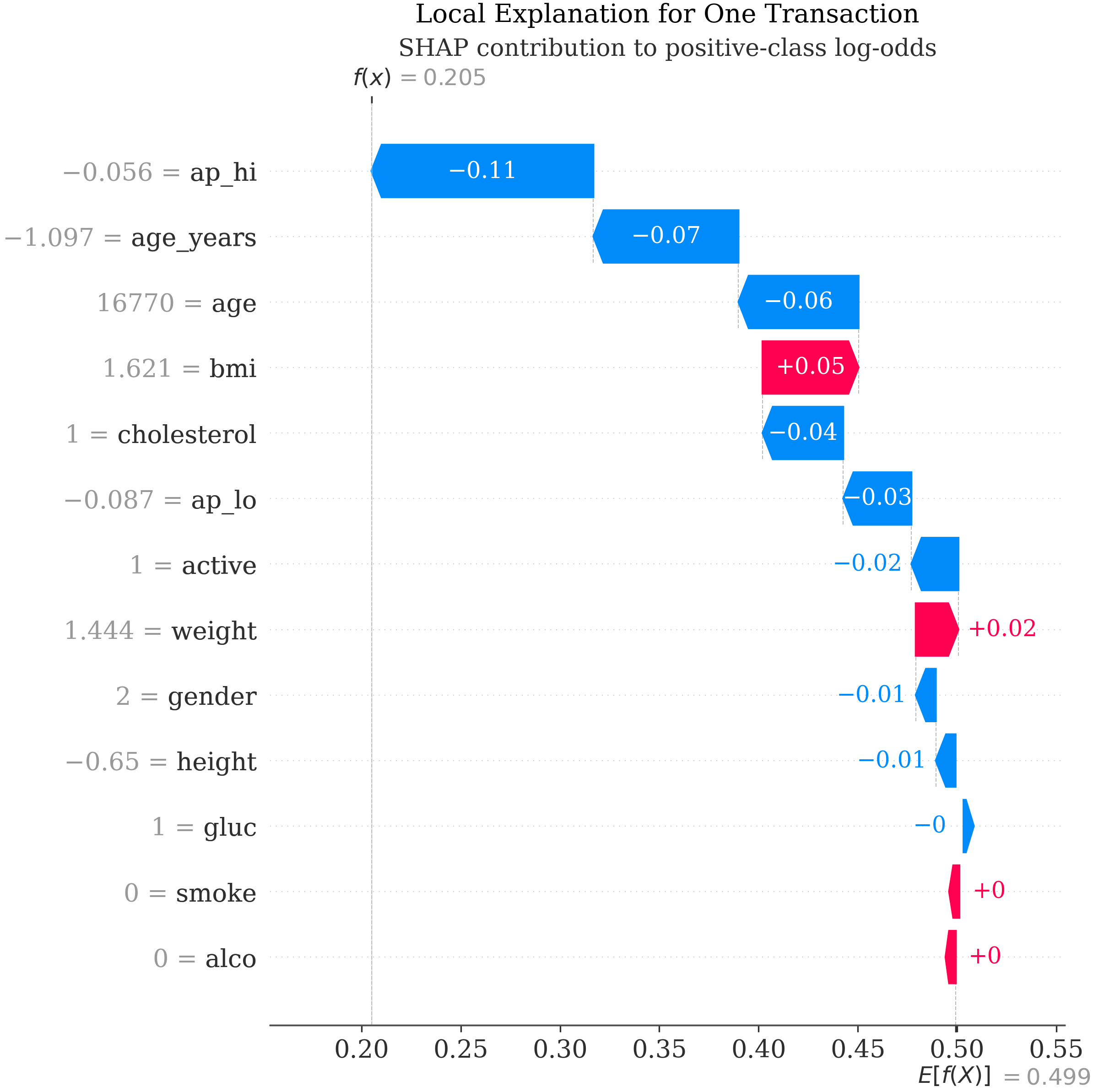}
\caption{Local SHAP waterfall plot for representative Kaggle test case.}
\label{fig:kaggle_local}
\end{figure}

Local explanations (Figure~\ref{fig:kaggle_local}) provided case-level interpretability, while clinician feedback (Table~\ref{tab:clinician-feedback}) rated SHAP global plots highest (4.5/5) for visual clarity and clinical utility.

\begin{table*}[!ht]
\centering
\caption{Clinician Evaluation of Explanation Methods (n=10)}
\label{tab:clinician-feedback}
\begin{tabularx}{\linewidth}{@{}XcccX@{}}
\toprule
\textbf{Explanation Method} & \textbf{Mean Rating (1--5)} & \textbf{SD} & \textbf{Representative Clinician Feedback} \\
\midrule
SHAP Global Plot & 4.5 & 0.5 & ``Clearly highlights the most important risk drivers without requiring statistical expertise.'' \\
Counterfactual Explanations & 4.2 & 0.6 & ``Provides actionable guidance for patient counseling.'' \\
LIME Local Explanations & 3.1 & 0.8 & ``Too technical for routine clinical use without additional support.'' \\
\bottomrule
\end{tabularx}
\end{table*}

\subsubsection{Clinician Usability Study Details}
\label{sec:clinician-study}

The clinician evaluation was conducted as an exploratory pilot study to assess the practical utility of different explanation modalities. Ten clinicians (cardiologists and internal medicine physicians) with 3--15 years of experience (mean${=}8.2$ years, SD${=}4.1$) participated in structured 30-minute sessions.

\textbf{Study Protocol:} Each clinician reviewed three anonymized patient cases presented with (1)~SHAP global summary plots, (2)~SHAP local waterfall plots, (3)~LIME local explanations, and (4)~counterfactual statements (e.g., ``Reducing cholesterol by 40~mg/dL would decrease risk by 15\%''). Participants rated each explanation type on a 5-point Likert scale (1$=$not useful, 5$=$very useful) and provided qualitative feedback via semi-structured interviews.

\textbf{Inter-rater Agreement:} Krippendorff's $\alpha$ for ratings was 0.71, indicating substantial agreement among clinicians despite the small sample size.

\textbf{Qualitative Themes:} Three key themes emerged:
\begin{itemize}
\item \textit{Visual Clarity:} SHAP global plots were praised for ``immediately showing what matters most'' without requiring statistical expertise.
\item \textit{Actionability:} Counterfactuals received positive feedback for supporting patient counseling (``I can tell the patient what to change'').
\item \textit{Technical Barrier:} LIME's perturbation-based explanations were critiqued as ``too abstract for clinical decisions without additional training.''
\end{itemize}

\subsection{MLOps Deployment and Real-Time Monitoring}

Our implementation integrates governance checks as mandatory deployment gates, providing systematic versioning of fairness audits, explanation artifacts, and performance metrics. Production monitoring revealed stable system operation across three cycles, with fairness metrics (DPD=0.07, EO=0.03) maintaining compliance without triggering alerts, ensuring operational governance.

\subsection{Computational Efficiency Analysis}

The computational costs of explainability procedures were substantial, while fairness auditing and drift detection proved resource-efficient (Table~\ref{tab:resource-consumption}). For scalable deployment, we implemented strategic sampling with TreeSHAP, balancing practical latency and explanation quality without sacrificing interpretive accuracy.

\begin{table}[ht]
\centering
\caption{Computational resource consumption by pipeline component}
\label{tab:resource-consumption}
\begin{tabularx}{\linewidth}{@{}Xcc@{}}
\toprule
\textbf{Pipeline Component} & \textbf{Time (s)} & \textbf{CPU (\%)} \\
\midrule
Bias Audit        & 41  & 22 \\
SHAP Explanation  & 125 & 61 \\
Drift Detection   & 33  & 19 \\
\bottomrule
\end{tabularx}
\end{table}

\section{Discussion}

This paper outlines how procedural justice, explicability, and operational governance can effectively converge in the same machine-learning pipeline. The suggested framework supports the needs of responsible artificial intelligence through automated fairness and model interpretability, ensuring a significant fairness benefit, and no harm to predictive power. Specifically, the demographic-parity gap was minimized to 0.04 (as compared to 0.31) and obtained positive usability feedback among the clinical stakeholders due to its improved transparency capabilities. Scalability was also justified by being applied to a large cardiovascular cohort of 70,000 cases, where it was found that the model achieved the same performance as the sample size increased. An RF model had an area under the curve of 0.776 (with strong fairness metrics (DPD= 0.012, EO= 0.027)) and an XGBoost model had an AUC of 0.800 (DPD= 0.021, EO= 0.021); both models met the pre-determined deployment thresholds. The optimized SHAP implementations that remained efficient to the system prevented computational overhead.

\subsection{Comparative study with existing methods}

Unlike the traditional toolkits, our approach condones justice through automated continuous-integration / continuous-deployment gates, as opposed to manual audit processes. In contrast to IBM AI Fairness 360 and Google What- If Tool, which only produce reports to be reviewed by humans, our framework performs a programmatic block of the implementation of models violating predetermined fairness criteria and automatically triggers retraining processes. On the same note, although MLOps platforms (MLflow and Kubeflow) provide solid infrastructure, they do not emphasize fairness as a mandatory deployment requirement but as optional metadata. The fact that fairness auditing requires only a relatively small amount of computational overhead, at 41 seconds per model run, a value that is less than 5\% of the most common training times. This working efficiency, versioned explanation artifacts, and strategic SHAP sampling help to enable reproducible model audits and meet operational latency limits.

\subsection{Scalability and clinical utility}

The large-scale replication experiment supports the idea that our fairness and explainability model can be introduced in the same way when using different sizes of datasets. This approach of obtaining global SHAP values on stratified samples and being able to explain results on a case-by-case basis is a good tradeoff between transparency requirements and computational expediency. The analysis of decision curves shows that the mitigation of fairness did not reduce clinical utility in the range of 10-20\%. The relative performance of mitigated and baseline models confirms the feasibility of deploying AI ethically with fairness standards set within reasonable boundaries in a manner that ethical AI concerns can be applied in a practical way without impairing its predictive capabilities.

\section*{Threats to Validity}

Various limitations worth noting are recognized in this work. Our inquiry focused primarily on gender fairness through demographic parity and equalized odds, excluding other salient fairness dimensions and intersectional characteristics that should be examined in the future. The use of U.S.-based datasets may limit extrapolativeness to dissimilar healthcare systems and populations; however, validation on a larger Kaggle cohort partially alleviates this issue. The fairness thresholds chosen represent pragmatic design choices that have real and practical impact on real-world outcomes. Though operationally indispensable, the calibration of such thresholds acknowledges the tension between endorsing fairness ideals and scaling to pragmatic deployment limitations. The computational intensity of SHAP is challenging for real-time clinical applications. Further optimization is necessary before these explanation methods become applicable in latency-sensitive environments. While our CI/CD pipeline demonstrates production readiness, its long-term reliability needs validation under evolving data conditions. Maintaining sustained performance in dynamic clinical settings is essential for building operational trust.

\section{Conclusion}
This paper demonstrates that integrating automated enforcement mechanisms into MLOps pipelines provides a feasible approach for incorporating fairness and explainability directly into production systems. The application of a reweighting algorithm significantly improved equity, as evidenced by the reduction in demographic parity difference from 0.31 to 0.04, a statistically significant improvement. Crucially, this fairness enhancement was achieved without consequential reduction in predictive accuracy, suggesting that trade-offs in clinical risk-prediction settings can be effectively managed. The framework's robustness is further supported by cross-dataset validation, which verified stable performance across diverse patient cohorts without requiring parameter retuning.

For the practitioners, this research translates ethical principles into practical engineering blueprints. By utilizing widely available development tools, our reference implementation provides actionable methodologies for establishing continuous integration gates for fairness, versioning model explanations, and automating retraining workflows. This approach empowers organizations to align ethical safeguards with their specific infrastructure, facilitating the transition from abstract aspirations to enforceable operational practices.

Despite these advances, several frontiers remain for future work. The scope of our clinician assessment, while informative, requires expansion beyond its current limited sample. Furthermore, our fairness examination should evolve to account for intersectional identities and move beyond binary gender analysis. Technically, reducing the computational cost of explanation methods like SHAP remains a priority for latency-sensitive deployments. While this framework establishes a foundational scaffold for scalable ethical AI amid increasing regulatory attention, the challenge of adapting to non-stationary data streams endemic to healthcare environments remains critical.

\balance
\bibliographystyle{IEEEtran}
\bibliography{references}

\end{document}